\documentclass[preprint,12pt,3p]{elsarticle}

\usepackage{amssymb}
\usepackage{amsmath}

\begin{document}
\begin{frontmatter}

\title{General null Lagrangians, exact gauge functions and forces in 
Newtonian mechanics}

\author{Z. E. Musielak and T. B Watson}
\address{Department of Physics, University of Texas 
at Arlington, Arlington, TX 76019, USA \\}

\begin{abstract}
A method to construct general null Lagrangians and their exact gauge 
functions is developed. The functions are used to define classical forces 
independently from Newtonian dynamics.  It is shown that the forces
generated by the exact gauge functions allow for the conversion of the 
first Newton equation into the second one.  The presented approach 
gives new insights into the origin of forces in Classical Mechanics.
\end{abstract}

\end{frontmatter}

\section{Introduction}
  
In Newtonian mechanics, the three laws of dynamics are independent, and 
the second law, $F = m a$, is used to define the force $F$ [1].  Moreover, 
the second law is also used to define the inertial mass $m$, which is the 
proportionality factor between the force $F$ and the acceleration $a$
 produced by it [2].  Hence, "the unknown is defined by the unknown" [1]. 
A possible solution of this circular definition problem is proposed in this 
Letter, where a novel method to introduce forces to Newtonian dynamics 
is presented.  The method is also used to demonstrate that the Newton 
law of inertia can be converted into the second Newton law of dynamics, 
which removes their independence.  

The presented method uses gauge functions that are similar to those that 
appear when the Galilean invariance of Newton's law of inertia is established 
[3].  The invariance requires the equation to preserve its form when the 
transformations representing the Galilean group of the metric are performed  
[4].  Thus, the Galilean transformations introduce the Galilean gauge that 
makes the equation invariant but not its corresponding Lagrangian [3,5].  
Recently, the gauge functions were constructed for the Newtonian law of 
inertia and used to make the Lagrangian Galilean invariant [6].  
 
In this Letter, the previous results [6] are generalized by introducing the 
exact gauge functions that require additional conditions to be imposed on 
the action [7].  Using these functions, the corresponding null Lagrangians 
are derived [8,9,10] and the Galilean invariance of both the Newton law 
of inertia and its Lagrangian is established.  The main advantage of this 
generalization is that it allows for the introduction of a general time-dependent 
force to the first Newton equation, thus converting this equation into the 
second Newton equation and implicitly preserving its Galilean invariance.  
This is our main result that gives new insights into the origin of forces in 
Classical Mechanics (CM), and in addition solves the longstanding problem 
of circular definitions of forces and inertial masses in Newtonian dynamics [1,2].   

\section{Galilean gauge function}

The metric of Galilean space and time is invariant with respect to all transformations 
that form the Galilean group of the metric [4,5].  Let ($x$, $y$, $z$) be Cartesian 
coordinates associated with inertial frames of reference that move with respect to 
each other with constant velocity $v_o$, and let $t$ be time that is the same in all 
these frames.  According to Newton's law of inertia, motion of a classical body is  
rectilinear and uniform with respect to any inertial frame.  

In one-dimension (along $x$) the law of inertia is then given by 
\begin{equation}
\hat D x (t) \equiv {{d^2 x (t)} \over {dt^2}} = \ddot x (t) = 0\ ,
\label{S2eq1}
\end{equation}
where $x(t)$ is a dynamical variable of the body in one inertial frame. 
The Lagrangian for this equation can be written in its standard form as  
\begin{equation}
L_{s} [\dot x(t)] = {1  \over 2} c_o \left [ \dot x (t) \right ]^2\ ,
\label{S2eq2}
\end{equation}
where $c_o$ is an arbitrary constant that represents inertial mass in CM. 

Let $(x, t)$ and $(x^{\prime}, t^{\prime})$ be inertial frames moving with 
respect to each other with the velocity $v_o$ = const, and let the system's 
origins be the same at $t = t^{\prime} = t_o = o$.  Then the Galilean 
transformations between these two frames are: $x^{\prime} = x - v_o t$ 
and $t^{\prime} = t$.  Let the initial conditions be $u (t) \vert_{t = 0} = 
\dot x (t)\vert_{t = 0} = u_0$ and $x (t)\vert_{t = 0} = x_o$, so that 
the solution of  Eq. (\ref{S2eq1}) can be written as $x (t) = u_o t + x_o$
and $x^{\prime} (t) = (u_o - v_o) t + x_o$, where $x_o = x^{\prime}_o$.

Applying these Galilean transformations to Eqs (\ref{S2eq1}) and (\ref{S2eq2}),
we find that $\hat D x (t) = 0$ and $\hat D x^{\prime} (t) = 0$ are of the 
same form, thus, Galilean invariant.  However, $L_{s} [\dot x(t)] = 
L^{\prime}_{s} [\dot x^{\prime} (t)] + L^{\prime}_{Gs} [x^{\prime} (t)]$, 
where 
\begin{equation}
L^{\prime}_{Gs} [\dot x^{\prime} (t)] = c_o \left [ \dot x^{\prime} (t) 
+ {1 \over 2} v_o \right ] v_o\ .
\label{S2eq3}
\end{equation}
It is easy to verify that $L^{\prime}_{Gs} [\dot x^{\prime} (t)]$ is the null 
Lagrangian [2] and that its Galilean gauge function [4] is 
\begin{equation}
\Phi^{\prime}_{Gs} [x^{\prime} (t), t] = c_o \left [ x^{\prime} (t) + 
{1 \over 2} v_o t \right ] v_o\ .
\label{S2eq4}
\end{equation}
The fact that the Galilean gauge function can be removed and $L_{s} 
[\dot x (t)]$ can be made Galilean invariant was recently shown [6] 
using other gauge functions.  In the following, those previous results 
are generalized by introducing the exact gauge functions, which are 
later used to define forces in Newton's equations.

\section{Exact gauge function}

The gauge function previously constructed [6] to make the standard 
Lagrangian for the law of inertia to be Galilean invariant can be written 
as  
\begin{equation}
\Phi [x(t), t] = {1 \over 2} c_1 x^2 (t) + c_2  x (t) t + c_4 x + 
c_6 t\ .
\label{S3eq1}
\end{equation}
where $c_1$, $c_2$, $c_4$ and $c_6$ are constants to be determined.
After the Galilean transformations of $\Phi [x(t), t]$, the resulting $\Phi^{\prime} 
[x^{\prime} (t), t]$ is of the same form as Eq. (\ref{S3eq1}) and the remaining 
terms are used to remove $\Phi^{\prime}_{Gs} [x^{\prime} (t)]$ given by Eq. 
(\ref{S2eq4}).  This guarantees that the Lagrangians $L_{s} [\dot x(t)]$ and 
$L_{Gs} [\dot x(t)]$ are of the same form [6].  In the following, the same 
procedure is used but for the newly defined exact gauge function, which 
also allows introducing forces into Newtonian dynamics.   

We generalize $\Phi [x(t), t]$ by replacing the constants in Eq. (\ref{S3eq1}) 
by functions of the independent variable that are continuous and at least twice 
differentiable, and obtain
\begin{equation}
\Phi_{n} [x(t), t] = {1 \over 2} f_1 (t) x^2 (t) + f_2 (t) x (t) t + f_4 (t) x + 
f_6 (t) t\ ,
\label{S3eq2}
\end{equation}
where $f_1 (t)$, $f_2 (t)$, $f_4 (t)$ and $f_6 (t)$ are to be determined.
Note that each term with a different function $f (t)$ is a partial gauge function.

Since the total derivative of $\Phi_{n} [x(t), t]$ gives the null Lagrangian 
$L_{n} [\dot x(t), x(t), t]$, we may write the action as 
\[
A [x (t); t_e, t_o] = \int^{t_e}_{t_o} ( L_s + L_{n} ) dt = \int^{t_e}_{t_o} 
L_s dt + \int^{t_e}_{t_o} \left [ {{d \Phi_{n}} \over {dt}} \right ] dt 
\]
\begin{equation}
\hskip0.25in = \int^{t_e}_{t_o} L_s dt + [ \Phi_{n} (t_e) - \Phi_{n} (t_o)]\ , 
\label{S3eq3}
\end{equation}
where $t_o$ and $t_e$ denote the initial and final times.  The gauge 
functions at the end points are constants, so they do not conflict with 
Hamilton's Principle that requires $\delta A [x (t)] = 0$ but they do 
add the difference between these constants to the value of the action.  
In other words, the value of the action is affected by the values of the 
gauge functions at the end points.  

Let $\Phi_{n} [x(t), t]$ be the exact gauge function if, and only if,
the value of the action remains unchanged, which means that both 
constants are either zero or the difference between them is zero;
following [7], we assume the former condition and use it to 
determine the end values of the arbitrary functions in the exact 
gauge function.  Other conditions on these functions are imposed 
by the Galilean invariance, which is now considered.  

\section{Galilean invariance}

To establish Galilean invariance of the law of inertia (see Eq. 
\ref{S2eq1}) and its standard Lagrangian (see Eq. \ref{S2eq2}),
we perform the Galilean transformations of $\Phi_{n} [x(t), t]$ 
and obtain $\Phi^{\prime}_{n} [x^{\prime} (t), t]$, and the 
following Galilean gauge function
\begin{equation}
\Phi^{\prime}_{Gn} [x^{\prime} (t)] = f_1 (t) \left [ x^{\prime} 
(t) + {1 \over 2} v_o t \right ] v_o t + \left [ f_2 (t) + f_4 (t) \right ] 
v_o t\ .
\label{S4eq1}
\end{equation}
To make the Lagrangian $L_{s} [\dot x(t)]$ Galilean invariant, it is 
required that the total derivative of the Galilean gauge (see Eq. 
\ref{S2eq4}) and the Galilean gauge function is zero, which means 
that the extra terms resulting from the Galilean transformations are 
eliminated.  Thus, the condition for the Galilean invariance can be 
written as 
\begin{equation}
{d \over {dt}} \left ( \Phi^{\prime}_{Gs} [x^{\prime} (t)] + 
\Phi^{\prime}_{Gn} [x^{\prime} (t)] \right ) = 0\ ,
\label{S4eq2}
\end{equation}
or $\Phi^{\prime}_{Gs} [x^{\prime} (t)] + \Phi^{\prime}_{Gn} 
[x^{\prime} (t)]$ = c = const.   Using $x^{\prime} (t) = (u_o - 
v_o) t + x_o$, we obtain
\begin{equation}
f_2 (t) = f_1 (t) \left ( {{1} \over {2}} v_o - u_o \right )\ ,
\label{S4eq3}
\end{equation}
\begin{equation}
f_4 (t) = c_o \left ( {{1} \over {2}} v_o - u_o \right )
- f_1 (t) x_o\ ,
\label{S4eq4}
\end{equation}
and $c = c_o v_o x_o$.

Then, the exact gauge function is given by  
\begin{equation}
\Phi_{n} [x(t), t] = {1 \over 2} \left [ v_o t - x (t) \right ] f_1 (t) 
x (t) + c_o \left ( {{1} \over {2}} v_o - u_o \right ) x (t) + f_6 (t) t\ .
\label{S4eq5}
\end{equation}
Taking $t_o = 0$, we determine the conditions required that the 
functions $f_1 (t)$ and $f_6 (t)$ satisfy in order for $\Phi_{n} 
[x(t), t]$ to be is exact.  These conditions are
\begin{equation}
f_1 (0) = {{2 c_o} \over {x_o}} \left ( {{1} \over {2}} v_o - 
u_o \right )\ ,
\label{S4eq6}
\end{equation}
and 
\begin{equation}
f_1 (t_e) = {{2 u_o - v_o - 2 t_e f_6 (t)} \over {v_o t_e - x_e}}\ ,
\label{S4eq7}
\end{equation}
shows that the end value of $f_ 1 (t_e)$ depends also on 
$f_6 (t_e)$, which remains undetermined.  

Having obtained the exact gauge function $\Phi_{n} [x(t), t]$, we
find the following exact null Lagrangian 
\[
L_{n} [\dot x (t), x (t), t] = {1 \over 2} v_o \left [ \dot x (t) t + x (t) \right ] 
f_1 (t) - f_1 (t) \dot x (t) x (t) + {1 \over 2} \left [ v_o t - x (t) \right ] 
\dot f_1 (t) x(t) \]
\begin{equation}
\hskip0.25in + {1 \over 2} \left [ v_o - 2 u_o \right ] c_o \dot x (t) 
+ \left [ \dot f_6 (t) t + f_6 (t) \right ]\ .
\label{S4eq8}
\end{equation}
This null Lagrangian makes the standard Lagrangian given by 
Eq. (\ref{S2eq2}) the Galilean invariant, which means that 
$L_s [\dot x (t)]$ and $L_s^{\prime} [\dot x^{\prime} (t)]$
are of the same form, and that $L_n [\dot x (t), x (t), t]$ and 
$L_n^{\prime} [\dot x^{\prime} (t), x^{\prime} (t), t]$ have 
exactly the same form.

\section{Definition of forces}

The energy function $\mathcal {E}$ in CM [11,12] is given by 
\begin{equation}
\mathcal {E} = \dot x {{\partial L} \over {\partial \dot x}} - L\ , 
\label{S5eq1}
\end{equation}
where $L$ is a Lagrangian.  If $L = L_s [\dot x (t)]$, then 
$\mathcal {E}= E_{tot}$ = const, with $E_{tot}$ being the 
total energy; in the case considered here, $E_{tot} = E_{kin} 
= c_o \dot x^2 / 2$.  However, when $L [\dot x (t), x (t), t] 
= L_s [\dot x (t)] + L_n [\dot x (t), x (t), t]$, the energy 
function then becomes
\begin{equation}
\mathcal {E} = {1 \over 2} c_o \dot x^2 (t) + {1 \over 2} 
\dot f_1 (t) x^2 (t) - {1 \over 2} v_o \left [ f_1 (t) + \dot f_1 (t) t
\right ] x(t) - \left [ \dot f_6 (t) t + f_6 (t) \right ]\ .
\label{S5eq2}
\end{equation}
Since the first term on the RHS is equal to $L_s [\dot x (t)]$,
we denote the remaining terms as $L_r [\dot x (t), x (t), t]$
and define $L_{\mathcal {E}} [\dot x (t), x (t), t] = L_s [\dot 
x (t)] + L_r [\dot x (t), x (t), t]$.  In general, the Lagrangian 
$L_r [\dot x (t), x (t), t]$ is not the null Lagrangian but it may 
be converted to one if, and only if, $f_1 (t)$ is of a special
form.  To determine this form, we substitute $L_r [\dot x (t), 
x (t), t]$ into the Euler-Lagrange equation and find $f_1 (t) =
C_r \exp [ v_o / 2 [( u_o - v_o / 2) t + x_o ]$; because of 
its special for, we keep $f_1 (t)$ to be arbitrary.  

Substituting $L_{\mathcal {E}} [\dot x (t), x (t), t]$ into 
the Euler-Lagrange equation, the following equation of 
motion is obtained
\begin{equation}
\ddot x (t) = F [x (t), t]\ ,
\label{S5eq3}
\end{equation}
where
\begin{equation}
F [x (t), t] = \frac{1}{c_o} \dot f_1 (t) x (t) - \frac{v_o}{2 c_o} 
\left [ f_1 (t) + \dot f_1 (t) t
\right ]\ .
\label{S5eq4}
\end{equation}

This shows that the derived exact null Lagrangian can be used to 
introduce the force $F [x (t), t]$ into the law of inertia and convert 
it into the second Newton law of dynamics.  The resulting force is 
linear in the dependent variable and can be of any dependence on 
the independent variable since $f_1 (t)$ is an arbitrary function of 
$t$ as long as it remains continuous and differentiable.  Note that 
$x (t) \neq u_o t + x_o$ as the solution  of the inhomogeneous 
equation of motion is now different and depends on the form of 
$F [x (t), t]$.

Now, if $F [x (t), t]$ does not depend on $x (t)$, then the force 
$F (t)$ can be any function of time as long as it is continuous
and differentiable, with the additional requirement that $f_1 (0)$ 
satisfies Eq. (\ref{S4eq6}); the function $f_1 (t_e)$ is not restricted.  
All constants in this force are uniquely determined by the initial 
conditions. Because of the dependence of $F (t)$ on the dependent 
variable $x (t)$, the derived inhomogeneous equation of motion is 
not Galilean invariant; similarly, the Lagrangian $L_{\mathcal {E}} 
[\dot x (t), x (t), t]$ used to obtain this equation is also not Galilean 
invariant.  In order to make them Galilean invariant, it is required 
to repeat the procedure described in Section 3 and 4 of this Letter, 
which will not be done here.  

The main advantage of $f_1 (t)$ being an arbitrary function 
of time is that a time-dependent force that drives a dynamical 
system may be represented by this function.  In general, the 
exactness imposed on the gauge functions sets up the initial 
value of the force, or in other words, its amplitude (at $t = 0)$ 
through the given initial conditions.  This shows that not all 
time-dependent forces may satisfy these conditions.  This is  
the physical limitation on the force caused by the exactness. 
The effect of the exactness is even more prominently seen in 
a special case of $f_1 (t) = c_1$ = const [6]; in this case, the 
resulting force is constant and uniquely determined by the 
initial conditions. 

Our method of defining $F (t)$ allows us to convert the law 
of inertia into the second Newton law of dynamics.  Our results 
show that this conversion requires the Lagrangian formalism 
and Galilean invariance, and that it does not guarantee that the 
resulting inhomogeneous equation of motion remains Galilean 
invariant; however, if the Galilean invariance is imposed on 
$F (t)$ then its form is further constrained.

Finally, let us point out that our novel approach and obtained 
results allow us to relate the first and second laws of dynamics, 
and solve the problem of the circular definition of the inertial 
mass and forces in Newtonian mechanics [1,2]. 

\bigskip\noindent
{\bf Acknowledgments}
We are grateful to an anonymous referee for carefully reading  
our manuscript and providing many detailed comments and 
suggestions that helped us to significantly improved the 
revised version of this paper.  We also thank L. Vestel for 
checking our derivations and comments on the manuscript.


\end{document}